# Ising-like agent-based technology diffusion model: adoption patterns vs. seeding strategies

Carlos E. Laciana[ab,1] and Santiago L. Rovere[b]

[a]Facultad de Ciencias Fisicomatemáticas e Ingeniería, Universidad Católica Argentina, Avenida Alicia M. de Justo 1400, Ciudad Autónoma de Buenos Aires, C1107AAR, Argentina.

[b]Laboratorio de Modelación Matemática, Facultad de Ingeniería, Universidad de Buenos Aires, Avenida Las Heras 2214 Ciudad Autónoma de Buenos Aires, C1127AAR, Argentina.

**Abstract**

The well-known Ising model used in statistical physics was adapted to a social dynamics context to simulate the adoption of a technological innovation. The model explicitly combines (a) an individual's perception of the advantages of an innovation and (b) social influence from members of the decision-maker's social network. The micro-level adoption dynamics are embedded into an agent-based model that allows exploration of macro-level patterns of technology diffusion throughout systems with different configurations (number and distributions of early adopters, social network topologies). In the present work we carry out many numerical simulations. We find that when the gap between the individual´s perception of the options is high, the adoption speed increases if the dispersion of early adopters grows. Another test was based on changing the network topology by means of stochastic connections to a common opinion reference (hub), which resulted in an increment in the adoption speed. Finally, we performed a simulation of competition between options for both regular and small world networks.

*Keywords*: Collective decision; Technology adoption; Ising model; Early adopters; Networks.

[1]*Corresponding author*. Tel./Fax: (54-11) 4514 3016.
*E-mail address*: clacian@fi.uba.ar (Carlos Laciana).

# 1. Introduction

The model for statistical physics known as "Ising model" was originally developed by Ernst Ising in 1925 to explain phase transitions in ferromagnetic materials [1]. It has been recently used in the simulation of several social processes [2], such as collective opinion formation [3][4][5, 6] or adoption of new technologies [7].

The versatility of the Ising model lays on the fact that the interaction effects for any given object with its neighbors is considered proportional to the number of neighbors in each state. Those objects can be spins (in up or down states), individuals with political positions (A or B), adopters and non-adopters of a new technology, population members infected and not infected with a contagious disease, etc.

Social networks are the main channels for the interaction in social models[2].In order to adapt the Ising model to a social context, we must add them to the original model [8].Network nodes represent individuals and links represent the communication channels between them.

The topological characteristics of social networks have considerable influence on interaction dynamics – in this case, the diffusion of innovations. Different topologies have been discussed in the literature, such as the small world [9][10], scale-free [11][12][13], modular [14] and regular [7]. Any of these topologies maybe used with the Ising model in a straightforward way [15]. However, when physical proximity among nodes is important, a regular lattice provides a good approach. In our analyses, we will mostly consider regular two-dimensional lattice and in some cases, small worlds networks.

In the original Ising model, the change in the spin orientation occurs when a threshold is reached in the mean field of the node. In a similar way, a threshold of decision must be reached in order to change the decision-maker agent state.

In most models of technology adoption (Delre, Jager and Janssen 2006) there are two basic terms which determine the threshold of decision: (a) social influence from a decision-maker's social network; and (b) the individual perception of a decision-maker agent about the benefits (or utility) of the new option. These two factors are combined into an "effective utility" that reflect the effects of both individual utility and social influence. The comparison of "effective utilities" (i.e., the relative effective utility) of both options leads to the selection of one option or another. The relative weight of the individual perception and social influence depends on the choice type[16]. For example, in fashionable markets (clothes, electronic gadgets) social influence has a strong weight, whereas on other choices (e.g., groceries), social influences are weaker.

As in physical systems, initial conditions have a strong influence on the evolution of a social system. In this particular case, technology adoption patterns are sensitive to the distribution of initial adopters (referred to as “seeding”) in the network. This effect was illustrated by Libai et al. [17], who showed that marketing strategies leading to different spatial distributions of early adopters introduce differences in the speed of adoption of a new product. The same issue was addressed by Delre et al. [18], who explored diffusion patterns resulting from alternative (spread out or concentrated) distributions of early adopters. In this paper, we explore systematically the influence of spatial dispersion of early adopters on the subsequent adoption dynamics.

A close relation exists between the distribution of initial adopters and the take-off time of the new product. In another way, Delre et al. [18] concentrate on the targeting and the timing of the promotions in relation to the take-off. That is previous to the generation of the distribution of initial adopters. Moreover, they use an agent-based model with a slightly different decision algorithm(both theindividual perception and the social influence must reach different thresholds independently, while in our approach the decision results from the effective utilities associated to each option). That decision is made in each time step, determining which option is adopted. Therefore the possibility of disadoption is introduced. This mechanism is useful when the adoption of a new product does not imply any investment (learning, technology or any other resource). If the last assumption is not satisfied, a model with no disadoption would be more appropriate. As our approach allows disadoption, it can be considered as a simple competition process.

In reference [7], the basic micro-structure of two or three initial adopters necessary to keep up diffusion was studied. In the present paper, we propose an extension by introducing many distributions of initial adopters with different dispersion degrees, in order to understand how the clustering of initial adopters affects the adoption speed.

In this paper, the simulations were performed using an agent-based model. Agent-based modeling is a way of doing thought experiments, obtaining, in many cases, non-obvious results and emergent patterns of the system [19]. The originality of this work does not lay in the introduction of a new statistical model (since the well-known Ising model has already been studied), but in the analysis of emerging evolutionary patterns associated to the adoption of a new product.

The paper is organized as follows. Section 2 presents a brief review of the Ising model and its application to modeling of innovation adoption in a social context. In Section 3 our specific implementation of the agent-based model is presented. Section 4 involves various experiments of technology diffusion including the study of adoption rate due to seeding effects, changes in the individual preference (both in space and time) and connection to a hub. Section 5 presents conclusions.

## 2. A model of technology diffusion

### 2.1 The Ising model in the physical context

The Ising model was originally developed to explain phase transitions in ferromagnetic materials. For example, suppose we are interested on describing a phase transition process in a ferromagnetic material. We can envision the material as constituted by a lattice of micro-magnets called "spins" that can interact with their nearest neighbors and with an external field. We will identify the state of spin in the $i^{\text{th}}$ position of the lattice by discrete variable $s_i$ that can take the values +1 or -1. If the system is constituted by N spins, its total energy is

$$E = \sum_{i=1}^{N} E_i = -\sum_{i=1}^{N} \left( \sum_{k=1}^{N} w_{ik} s_k + h \right) s_i \equiv -\sum_{i=1}^{N} m_i s_i \quad , \tag{1}$$

Where $w_{ik}$ is the coupling strength between nearest neighbor spins, and $h$ is a constant external magnetic field. $E_i$ is the energy associated with spin$i$, where $m_i$ is the magnetic field around spin $i$. From Eq. (1) it follows that

$$E_i(s_i = \pm 1) = \mp m_i \, .$$

The probability of finding spin $i$ in state $s_i \in \{+1;-1\}$ is given by the Boltzmann-Gibbs distribution:

$$P(s_i = \pm 1) = \frac{1}{1 + e^{\mp 2\beta m_i}} \quad , \tag{2}$$

With $\beta = 1/kT$, where $k$ is Boltzmann´s constant, and T is the system temperature. Moreover, using detailed balance condition, that ensures convergence to equilibrium, Eq. (2) can be used to calculate the transition probability between the two states.In the following section, an interpretation for each term of Eq. 2 will be given in the context of a social system.

## 2.2 Using the Ising model to simulate the diffusion of innovations

The original Ising model can be adapted to a social context in order to simulate the adoption (or disadoption) of an innovation. In a social system, the spins in the Ising model can be interpreted as N individuals, households or firms – hereafter referred to as "agents" – who must choose between two options: A ($s_i = +1$) or B ($s_i = -1$); options A and B may represent, for example, new and existing technologies respectively. The magnitude $m_i$ can be interpreted as the "relative effective utility" between options A and B. In order to use a more familiar notation, in the social context we will denote the relative utility as $\Delta U_i$ instead of $m_i$. As in Ref. [7] and by analogy with Eq. (1), we can write $\Delta U_i$as:

$$\Delta U_i = 2\alpha_i \sum_{k=1(k\neq i)}^{N} \left( \frac{J_{ik}}{N_v^i} \right) s_k + 2(1-\alpha_i)\Delta u_i \qquad (3)$$

The r.h.s. of Eq. (3) has two terms. The first one (analogous to the interaction term between spins in the original Ising model) describes the contribution of social influence from decision-maker $i$'s social network, whereas the second term (analogous to the external magnetic field in the physical model) describes the contribution of $i$'s individual preference for options A or B, irrespective of other agents [3, 6, 20].

$N$ is the total number of agents, $N_v^i$ the number of individuals connected to agent$i$ by a first-order link in the social network (i.e., those agents assumed to havea social influence on $i$'s decisions),$J_{ik}$quantifies the social influence of agent $k$ on agent $i$'s decisions (in all simulations we assume $J_{ik} = 1\ \forall i,k$, that is, all agents have the same influence over other agents). Factor$\alpha_i \in [0, 1]$ weighs the relative importance of social influence and individual preference on the overall utility of a given option/product: if $\alpha_i$>0.5,social influence is more important than the individual preferenceand vice versa. In all simulations we will use $\alpha = 0.5$ (i.e., both components of effective utility are assumed to have the same weight).

The individual preference component reflects an agent's idiosyncratic preferences for options A or B, irrespective of other agents [3, 6, 20]. Individual preference $\Delta u_i$ in Eq. 3 can be defined as

$$\Delta u_i = \frac{u_i(\mathrm{A}) - u_i(\mathrm{B})}{\max[u_i(\mathrm{A}); u_i(\mathrm{B})]}, \tag{4}$$

Where $u_i(\mathrm{A})$ and $u_i(\mathrm{B})$ represent the utilities experienced by an individual if he chooses option A or B respectively.Here, function $u(x)$ reflects a broad measure of desirability, even including non-economic factors, and can take different forms: it may represent, for example, the expected value of economic profits from a given option. Values of $\Delta u_i$ in Eq. 4 are dimensionless and range within interval [-1; 1].

Note that many quantities in Eq. 3 are indexed by agent. Unless otherwise specified, all our simulations assume that these quantities are the same for all agents. That is, possible different personal characteristics of each agent (e.g., risk aversion that may influence valuation of a given option) are not taken into account. This is analogous to considering a set of agents with a mean value for each parameter.

## 2.3 Decision algorithm

The original Ising model assumes an equilibrium "temperature" that defines the probability of permanence in each spin state. In a social context, inclusion of a system "temperature" introduces global uncertainty in a decision (affecting both social and personal components), turning it into a stochastic event. In such context, the temperature T (Eq. 2) can be interpreted as random noise, due to erratic circumstances that influence the opinion of *all* agents about the advantages of selecting one of the two options [21, 22]. For example, if the agents are farmers deciding on adoption of a new crop variety, temperature may represent fluctuations related to *"...epidemics, annual weather fluctuations, political events. These events change the perception of farmers and might make them take decisions that they would not have taken under 'normal' circumstances"*[7].

The effects of temperature on the probability of occurrence of a given event (adoption or non-adoption) will not be analyzed here; instead, in all subsequent simulations we only consider the case of T=0. From Eq. (2) it follows that, when T = 0 (i.e., no random noise), the probability of a given event is fully determined by the sign of $\Delta U_i$:

$$\lim_{T\to 0} P(s_i = +1) = \begin{cases} 1 & if \quad \Delta U_i > 0 \\ \frac{1}{2} & if \quad \Delta U_i = 0 \\ 0 & if \quad \Delta U_i < 0 \end{cases} \tag{5}$$

and conversely

$$\lim_{T\to 0} P(s_i = -1) = 1 - \lim_{T\to 0} P(s_i = +1)\,. \tag{6}$$

In this limit, the model is deterministic except for $\Delta U_i = 0$, when both events have equal chance.

## 3. Agent-based model of technology diffusion

Many recent studies of diffusion and adoption rely on agent-based modeling, a powerful simulation technique that is very promising for developing new diffusion theory[19, 23, 24]. Zenobia et al. [25] assess the strengths, opportunities, weaknesses, and threats facing agent-based modeling in technological innovation research. An agent-based model (ABM) consists of a collection of autonomous decision-making entities ("agents"), an environment, and rules that determine sequencing of actions in the model. Each agent has sensory capabilities and makes decisions on the basis of a set of rules[26]. Agents can interact either indirectly through a shared environment and/or directly with each other through markets and, especially, through social networks [27, 28].

We implement here an agent-based model of technology diffusion based on the Ising model described in the previous section. Many software frameworks are available that facilitate ABM development; we use REPAST Simphony, an open-source framework maintained by Argonne National Laboratory[29].

The environment of the ABM is a 2-D lattice with 10,000 nodes(arranged in a 100 X 100 grid). Each node represents an agent that makes decisions about the adoption (or disadoption) of a technology.Agents located on the edges of the lattice have fewer neighbors because we do not assume periodic boundaries (i.e., a torus).

At each time step of the simulation, an agent decides if he/she adopts a technological innovation based on the overall relative utility of the new technology (Eq. 3). The decision algorithm is given in Eq. 5. That is, the sign of overall relative utility $\Delta U_i$ determines if the agent goes from state -1 to +1 (adoption) or viceversa (disadoption) or, alternatively, stays in his current state. Each simulation continues until only small fluctuations are observed in the adoption pattern, or a given option prevails completely.

## 3.1 Adoption threshold

As shown in a previous section, the adoption of an innovation by agent *i*at time *t* depends on the sign of the overall relative utility $\Delta U_i$ of old and new products or technologies. It follows that, for each value of the individual preference between options ($\Delta u_i$ in Eq. 3), there is a threshold of social influence for agent *i* (expressed as the number of adopter neighbors) below which adoption does not take place. We denote the number of neighbors of agent *i* in states $s_i = +1$ and $s_i = -1$ as $v_{i+}$ and $v_{i-}$, respectively. It follows from Eq. (3) that

$$\sum_{k=1(k\neq i)}^{N} J_{ik} s_k = v_{i+} - v_{i-} \tag{7}$$

The assumption used in this calculation is that all components $J_{ik}$ are zeroes or ones; this means that there are no agents with higher influence than others, i.e., all social links have the same weight. Therefore, the condition $\Delta U_i > 0$ for the adoption of option +1 by agent *i* (assuming $\alpha = 0.5$) can be rewritten more explicitly as:

$$\frac{1}{N_v^i}(v_{i+} - v_{i-}) + \Delta u_i > 0 \tag{8}$$

Then, taking into account that $v_{i+} + v_{i-} = N_v^i$, the condition for adoption is:

$$v_{i+} > \frac{1}{2} N_v^i (1 - \Delta u_i) \equiv v_{i+}^{\min} \tag{9}$$

where $v_{i+}^{\min}$ is the minimum number of first-order social contacts necessary for adoption.

# 4. Results

## 4.1 Effect of seeding on diffusion speed

Previous work has shown that technology adoption patterns are sensitive to the seeding or distribution of early adopters (i.e., those who already have adopted the innovation when simulations are started). For example, [17] showed that marketing strategies leading to different spatial distributions of early adopters can, in turn, introduce differences in the speed of adoption of a product. Similarly, [18] explored diffusion patterns resulting from spread-out or concentrated distributions of early adopters.

This section explores the effects of the spatial distribution of initial adopters on the speed of diffusion of a new technology. We simulate diffusion patterns resulting from increasing spatial dispersion in the distributions of a constant number (N = 100) of initial adopters. To define the positions of the 100 initial adopters on a 100 x 100 lattice, we draw random values from two Gaussian distributions representing, respectively, two uncorrelated variables: the x- and y-coordinates on the grid. The two distributions are centered on the grid. To quantify the spatial dispersion of adopters, we introduce an ad-hoc, computationally convenient definition of dispersion. By analogy with the univariate standard deviation, we define a square area, centered in the middle of the grid, that includes about 68% of the initial adopters (in this case, 68 agents), and define "dispersion" $\sigma$, as half the side of this square, measured in number of positions in the lattice (Fig. 1a).

[Insert Figure 1 about here]

We will consider the lattice space as the potential market for a new product or technology. In the beginning, the market is saturated with an earlier product which will be in competition with the new one. Possible marketing strategies for the new product are to concentrate advertising resources in a small area or, alternatively, target a broader area. The question that arises is: which of these strategies is more efficient?. To address this question, we will assume a direct correspondence between the pattern of advertising and the distribution of initial adopters. That is, if advertising is concentrated in a small region, early adopters will be concentrated in the targeted area. Conversely, early adopters will be more dispersed when the advertising campaign targets an extended area. The spatial dispersion of initial adopters is quantified by $\sigma$ (see discussion above).

[Insert Figure 2 about here]

Fig. 2 shows the association between saturation time (i.e., the time at which the space is completely occupied by the new product) and a broad range of values of spatial dispersion of early adopters $\sigma$ ranging from ≈4 (a very tight distribution) to ≈41 (which approximates a uniform distribution). Results are simulated for two alternative network topologies: (a) a regular network (ties are defined by first-order neighborhood in the lattice) and (b) a "small world" network generated using a rewiring procedure similar to that described by Watts and Strogatz[9]. The rewiring readjusts the edges for each node, moving an edge to another randomly selected node with probability p=0.005; this rewiring probability is inside the interval [0.003,0.02] that characterizes small-world networks in a 2-D network [30]. Simulations also are carried out for three different values of $\Delta u$ (0.8, 0.6 and 0.4), indicating respectively large, intermediate and small differences in the relative utility of the two products; as all $\Delta u$ values are positive, we assume that the innovation has advantages over the existing technology or product.

As $\Delta u$ decreases, adoption will occur only if social influence effects are higher: i.e., a larger number of adopter neighbors is needed. The three $\Delta u$ values used in simulations require 1, 2 and 3 adopting neighbors, respectively, as shown in Eq 9. For each combination of $\Delta u$ and network topology, 100 simulations are run with different distributions of early adopters; results plotted in Fig. 2 represent the average of the 100 runs.

For $\Delta u = 0.4$, saturation is reached in most experiments up to a certain dispersion of initial adopters ($\sigma$ values up to about 15). In this range, diffusion proceeds faster for lower $\sigma$ values (i.e., time to saturation is lower). This pattern is qualitatively similar for regular and small-world networks, although diffusion is faster for the latter (the offset between curves for both topologies seems fairly constant). As dispersion increases ($\sigma \geq 15$)most experiments do not lead to saturation, and therefore saturation time is infinite (thus, no line is shown). When the adoption threshold is lower ($\Delta u = 0.6$ or 0.8), saturation is reached in every experiment. Unlike the previously described situation, diffusion proceeds faster when the dispersion of initial adoption increases (as $\sigma$ grows). Small-world networks show faster diffusion than regular networks because of the existence of shortcuts (weak ties).

## 4.2 Changes in the spatial distribution of individual preference

In real situations, individual preferences for competing technological options can be different for each decision-maker. In some cases, individual preference may be a function of the spatial location of an agent. For example, suppose we are comparing two crop varieties, one of which has a higher tolerance to stresses associated with water shortages. In this case, it is likely that the drought-tolerant variety will have a higher utility in drier locations where water stresses are more likely, and lower utility in places where rainfall is plentiful.

To explore this effect, we perform simulations in which $\Delta u$ decreases radially from the center of the lattice. The value of $\Delta u$ at any node of the lattice is given by

$$\Delta u = \Delta u_0 e^{-\mu\left(\frac{d}{l}\right)^2} \tag{12}$$

where Δu = 0.8 is the utility at the centre of the lattice, $l$is a length scale associated with the grid (in this case, $l = \sqrt{N}/2$, where $N$is the total number of agents) and $\mu$is a dimensionless parameter that describes how quickly $\Delta u$ changes with distance (in this case μ = 3). We also carry out experiments preserving the $\Delta u$ gradient but varying the network topology through a rewiring process similar to the one described above. Both sets of experiments are based on 100 early adopters with a uniform spatial distribution.

From Equations 9 and 12, it can be inferred that there are concentric regions in the lattice with different numbers of adopter neighbors required for adoption. These adoption thresholds are indicated in Figs. 3a and b for neighborhoods of size 4 and 8 respectively.

[Insert Figure 3 about here]

Figure 3 displays adoption patterns after the diffusion dynamics have been completed (i.e., when no further changes occur). In most cases, total adoption does not occur (an exception is Fig. 3g). For regular lattices, the final distributions involve a central "island" of adopters with compact and regular geometric shapes: a square when four neighbors (a von Neumann neighborhood) are considered (Fig. 3c), and an octagon when eight neighbors (a Moore neighborhood)are used (Fig. 3d). These patterns can be easily interpreted. Given the radially decreasing $\Delta u$ pattern, as we move away from the center of the lattice, more adopter neighbors are required for adoption to occur. At a given distance from the center, three adopter neighbors are needed. This threshold, however, is difficult to reach given the scattered distribution of initial adopters therefore adoption does not proceed beyond this boundary.

For a regular lattice (rewiring probability $p = 0$), social influence involves only geographical vicinity therefore the "island" of adopters grows outwards regularly, in crystal-like fashion. In contrast, when randomness in ties is introduced through a rewiring probability $p > 0$, final adoption "islands" no longer have regular shapes. This is because neighbors no longer include only adjacent agents in the lattice. The edges of the adoption islands correspond to adopters who are farthest from the center of the space, and thus are most likely to become non-adopters if they lose an adopter neighbor due to rewiring. For $p = 0.25$, adoption islands are approximately circular, with a few adopters outside (Fig. 3e-f).

The size of the adoption island is larger for neighborhoods of size 8. The larger size is due to the region in this neighborhood where the adoption threshold is 3; this region does not exist for neighborhoods of size 4. To explain this behavior, one must remember that that social influence is defined by the proportion of adopters in the neighborhood, not their absolute number. For an 8-agent neighborhood, in the zone where the adoption threshold is 3 an agent needs 37.5 % of neighbors (3 out 8) to be adopters. In the same zone, for a 4-agent neighborhood an agent needs 50% of adopter neighbors (2 out of 4) to adopt.

When rewiring probability is increased from $p = 0.25$ to $p = 0.50$, the adoption patterns are very different for neighborhoods of size 4 and 8 (Figs. 3g and 3h). For neighborhoods of size 4, complete adoption is observed (Fig. 3g), that is, an increase in adoption with respect to the case in which $p = 0.25$ (Fig. 3e). In contrast, for an 8-agent neighborhood there is a lower number of adopters (Fig. 3f); furthermore the increase in rewiring probability results in a decrease in adoption for the same neighborhood (Fig. 3h vs 3f).

## 4.3 A particular change in the topology of the social network: connection to a "hub"

Diffusion of innovations is thought to be strongly influenced by people who have a large number of ties to other people [31]. In the social network literature, these individuals are referred to as hubs, influentials, opinion leaders, or "members of a royal family," watched by many others in the network[32].

In this section we study the dynamics of adoption when we introduce a hub agent connected to a large number of agents in the lattice. The literature suggests that hubs may have a strong influence on other agents' opinions or actions – thus the tie to the hub would have a stronger influence than other ties. Nevertheless, here we assign to the hub the same influence as any other agent, so we can observe the sensitivity of the model to the changed network topology.

A similar model is used in [17], where each individual is influenced by two groups of agents: his or her nearest neighbors ("neighborhood effect") and agents from other regions, denoted as "relatives" and connected through weaker ties. However, the Libai et al. [17]model differs slightly from the one used here: the term corresponding to individual preference for an innovation does not appear (i.e., only social influences are considered) and irreversibility of transition is assumed – consumers cannot disadopt once they have adopted. Our model, in contrast, is symmetrical (adoption and disadoption are both allowed), as in models used foropinion formation.

To perform simulations in this section, the original regular network was modified by stochastically selecting 100 agents;these agents were then connected to a hub – arbitrarily located at the center of the lattice – by replacing stochastically with a probability $p = 0.1$ one of their incoming links from a neighbor. Although individuals with many social ties are not necessarily innovators [33], our simulated hub is initialized as an adopter.

[Insert Figure 4 about here]

Figures 4a and 4d display the proportion of adopters as a function of time for experiments with and without a hub, and for $\Delta u = 0.6$ and $\Delta u = 0.4$, respectively. For both $\Delta u$ values, adoption proceeds faster when a hub is present. Figures 4b, c, e and f show snapshots of adoption patterns for comparable stages of the diffusion. These plots confirm the faster spread of an innovation when a hub is present, as suggested by the much larger number of adopters at the same step of a simulation. Furthermore, the growth of adoption "islands" is different with and without the hub. When there is no hub and only local neighborhood influences diffusion, patterns show a crystal-like growth as shown previously. The hub introduces a random component in the network topology, and adoption can spread in any direction.

## 4.4 A simple model of competition between two options

In all previous experiments, the relative utilities of old and new technologies or products were defined at the beginning of an experiment and did not change throughout the simulation. In this section, in contrast,we simulate dynamic competition between two products by allowing changes in $\Delta u$ through out a simulation.

We assume that a new product is introduced into a market and starts competing with a pre-existent product. At its introduction, the new product is better than the older one (i.e.,$\Delta u > 0$) and thus gains market share. However, once the manufacturers of the older product notice that the new product has reached a certain proportion of the market (which we denote as the "critical market share" or CMS), they react by introducing improvements such that the older product matches the utility of the new product (i.e., $\Delta u$ becomes zero).We explore how the dynamics of competition evolve in response to this change in relative utilities.

First, we study the probability that the new product will continue to prevail (i.e., maintain or increase its market share) after the enhancement of the older product. We denote the probability of prevalence as $\zeta$, and we estimate it as the relative frequency with which the new product continues to prevail in a set of 100 simulations for a given set of conditions.

[Insert Figures 5 (a) and 5 (b) about here]

Figure 5a shows how $\zeta$ varies as a function of CMS for a regular network. When CMS $< 0.5$, there is a higher probability of prevalence when the seeding of initial adopters is spatially disperse.In contrast, when CMS $\geq 0.5$, concentrated distributions of initial adopters have a higher probability of prevalence. This transition happens more abruptly when the early adopters are more concentrated because in this case the amount of non-equivalent configuration is smaller. Figure 5b shows results for a small-world network (with a rewiring probability of 0.005). In this case, changes in the probability of prevalence are much less sensitive to the initial dispersion of early adopters. In this case, the probability of prevalence evolves much more regularly as a function of CMS: the higher the share gained by a new, superior product before it is matched by a previous product, the higher are the chances that the new product will retain or enhance the market share during the period when it was better than the older product.

## 5. Conclusions

The main conclusions of our experiments are as follows:

- When a new product or technology has clear advantages over existing products (i.e., $\Delta u > 0.4$), the innovation is adopted more quickly when early adopters are spatially disperse. In contrast, when the new product is only slightly better ($\Delta u = 0.4$) than the existing option, market saturation is not reached by the new product if early adopters are dispersed. Therefore, in the case of new products without clear advantages, marketing strategies should aim to develop a concentrated set of early adopters. In contrast, when the new product is clearly superior, the best marketing strategy seems to be to attain a broader spatial distribution of early adopters. These conclusions are applicable to both regular and small-world networks.

- We explored a situation in which the advantages of an innovation ($\Delta u$) decrease regularly as a function of distance from the center of the lattice. Adoption thresholds have a regular pattern of concentric circles. However, the functional form of $\Delta u$ constrains most adopters to a compact central region. Therefore, equilibrium adoption patterns look like "islands of adoption" (octagonal for neighborhoods of size 8, square for size 4 neighborhoods). These symmetrical patterns are reached independently of the degree of dispersion of the initial adopters. The symmetrical patterns are broken when other topologies are used.

- The presence of agents who have a large number of ties to other agents (referred to as hubs, influentials, or opinion leaders) accelerates the adoption of a new technology or product. Moreover, geometric patterns in the diffusion of a new product are observed, which are very different to those obtained when only spatial neighbors are considered.

- Finally, we performed an experiment in which the initial advantage of a new product is subsequently matched by enhancements in the older competing product. For a regular network with spatially concentrated early adopters, a marketing strategy should aim to achieve quickly at least half of the market share. In this case, reaction by the competition does not decrease the market penetration initially gained by the new product during the period when it was superior to the alternative. If the distribution of initial adopters is dispersed, the chances of the new product retaining its market share decrease. For small-world networks, spatial distribution of early adopters does not influence significantly the probability of retaining market share.

## Acknowledgements

Useful comments by two anonymous reviewers are gratefully acknowledged. This research was supported by two U.S. National Science Foundation (NSF) Coupled Natural and Human Systems grants (0410348 and 0709681). Additional support for one of the authors (S. R.) was provided by grant CRN-2031 from the Inter-American Institute for Global Change Research (IAI), which is funded by NSF Grant GEO-0452325, and by the University of Buenos Aires.

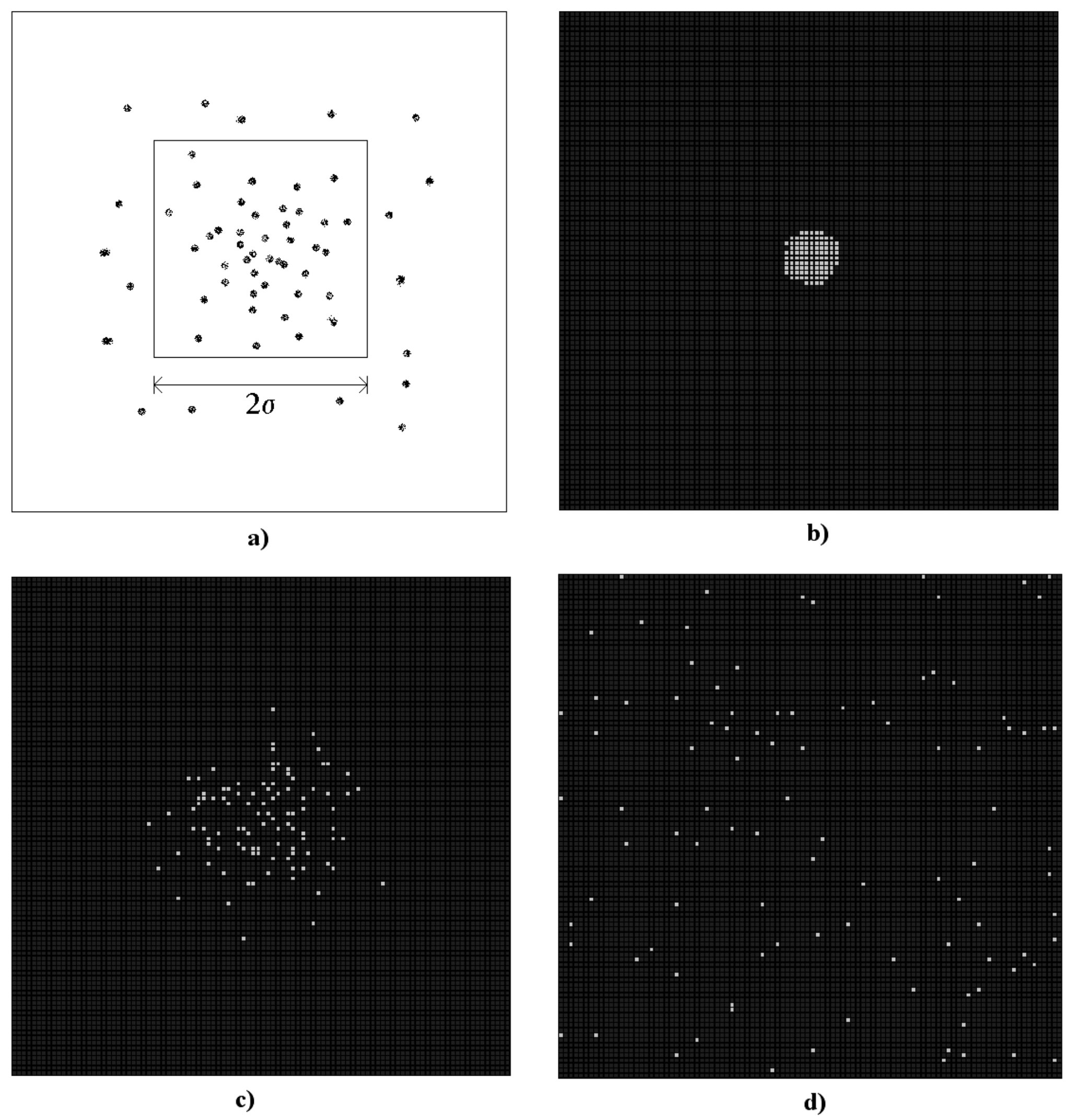

Figure 1. (a) Definition of the ad-hoc metric $\sigma$of spatial dispersion of early adopters. a) $\sigma$is defined in a square region of 2σ x 2σ such as about 68.2% of early adopters are included there. b) Initial distributionfor σ ≈ 4. c) Initial distribution for σ ≈ 13.5. d) Initial distribution for a uniform distribution

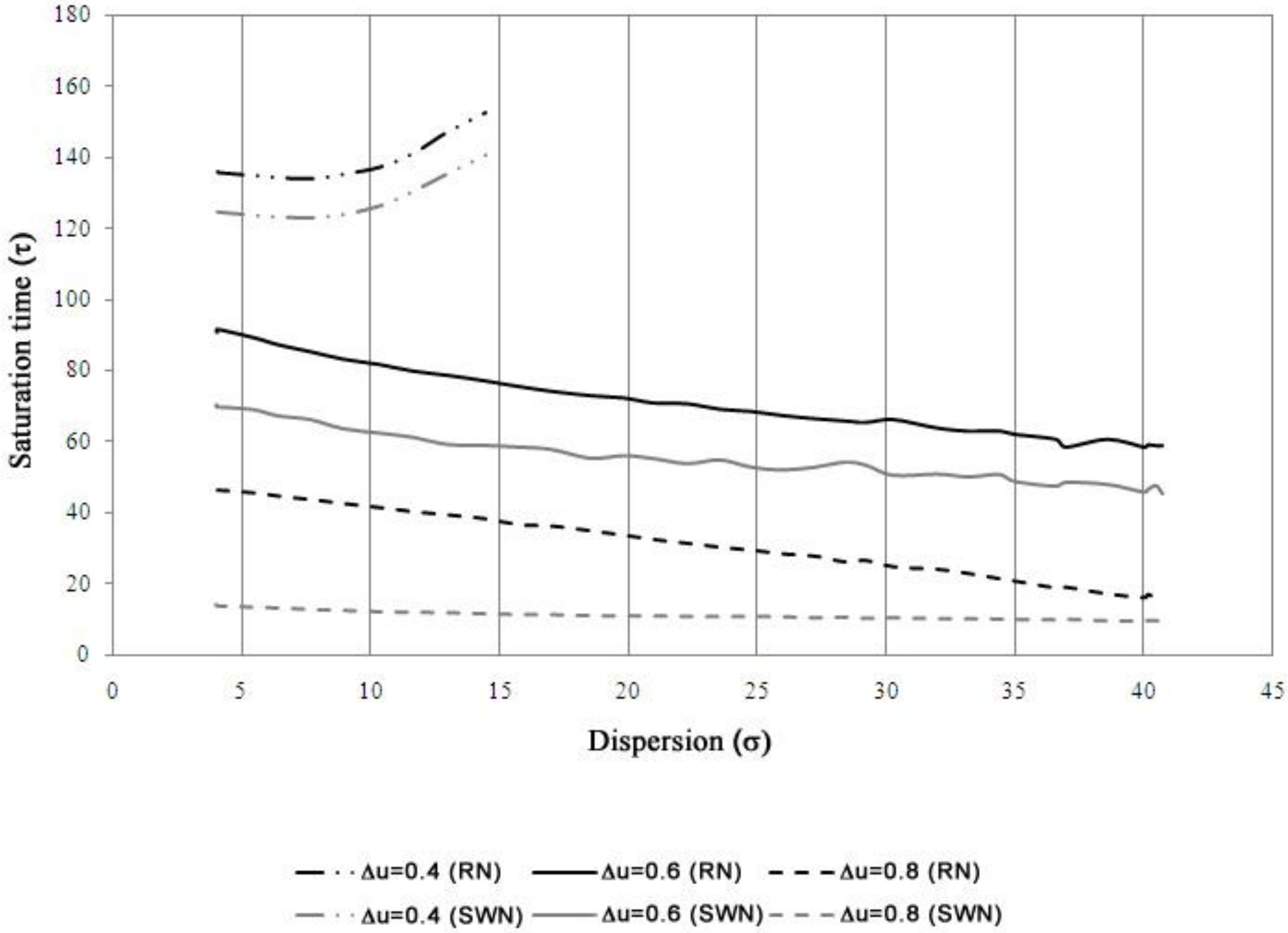

Figure 2: Saturation time ( $\tau$ ) versus dispersion parameter ( $\sigma$ ) in the distribution of early adopters for 2 different values of $\Delta u$. Black lines represent results for regular networks (RN) and grey lines, results for small world networks (SWN).

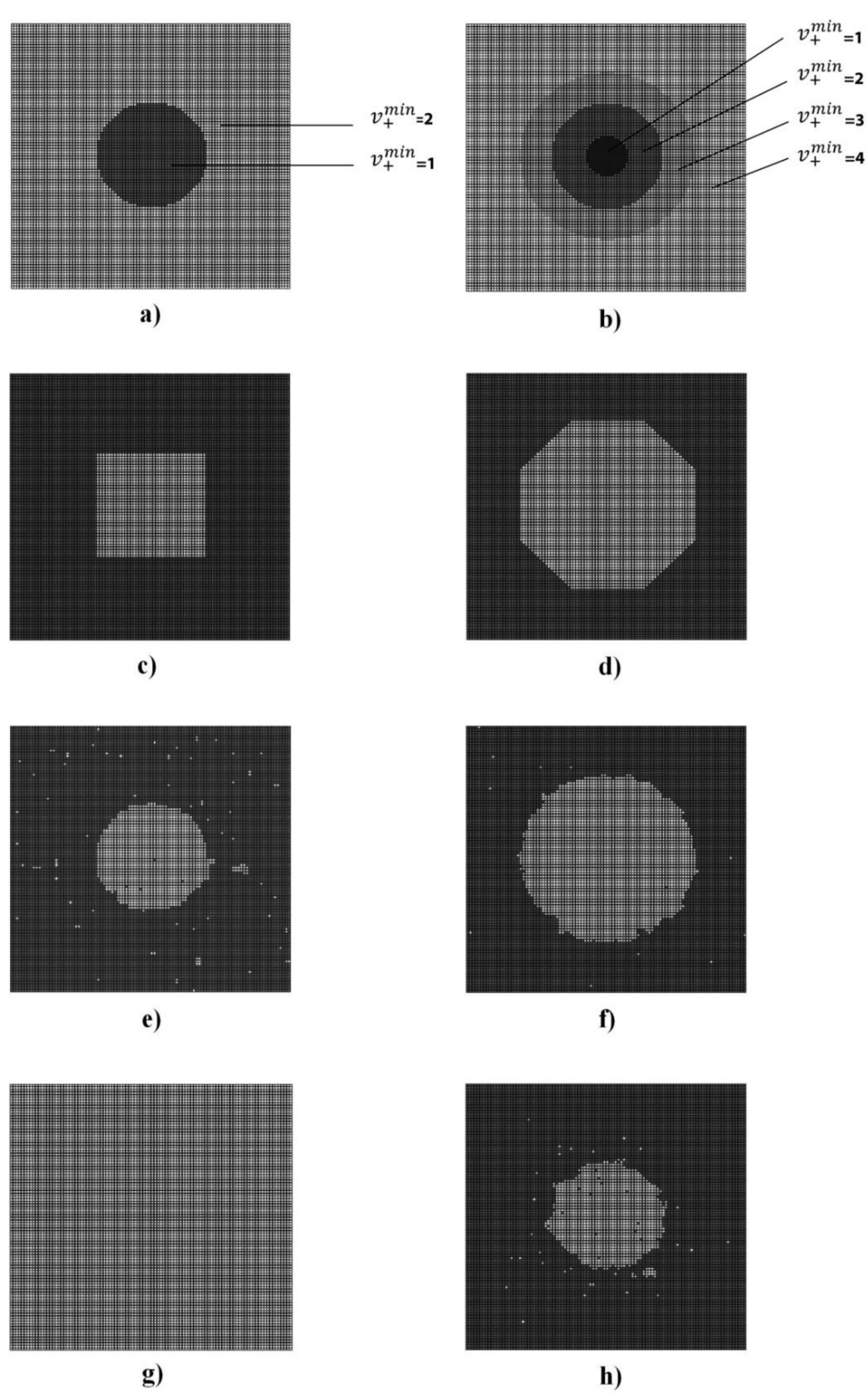

Figure 3: Adoption patterns associated to changes in the topology; a) and b) shows the adoption thresholds $v_{i+}^{min}$ for a neighborhood of 4 and 8 agents respectively; c) and d) the islands of adopters at the end of the process, with rewiring probability p = 0; e) and f) with p = 0.25 and g) and h) with p = 0.5.

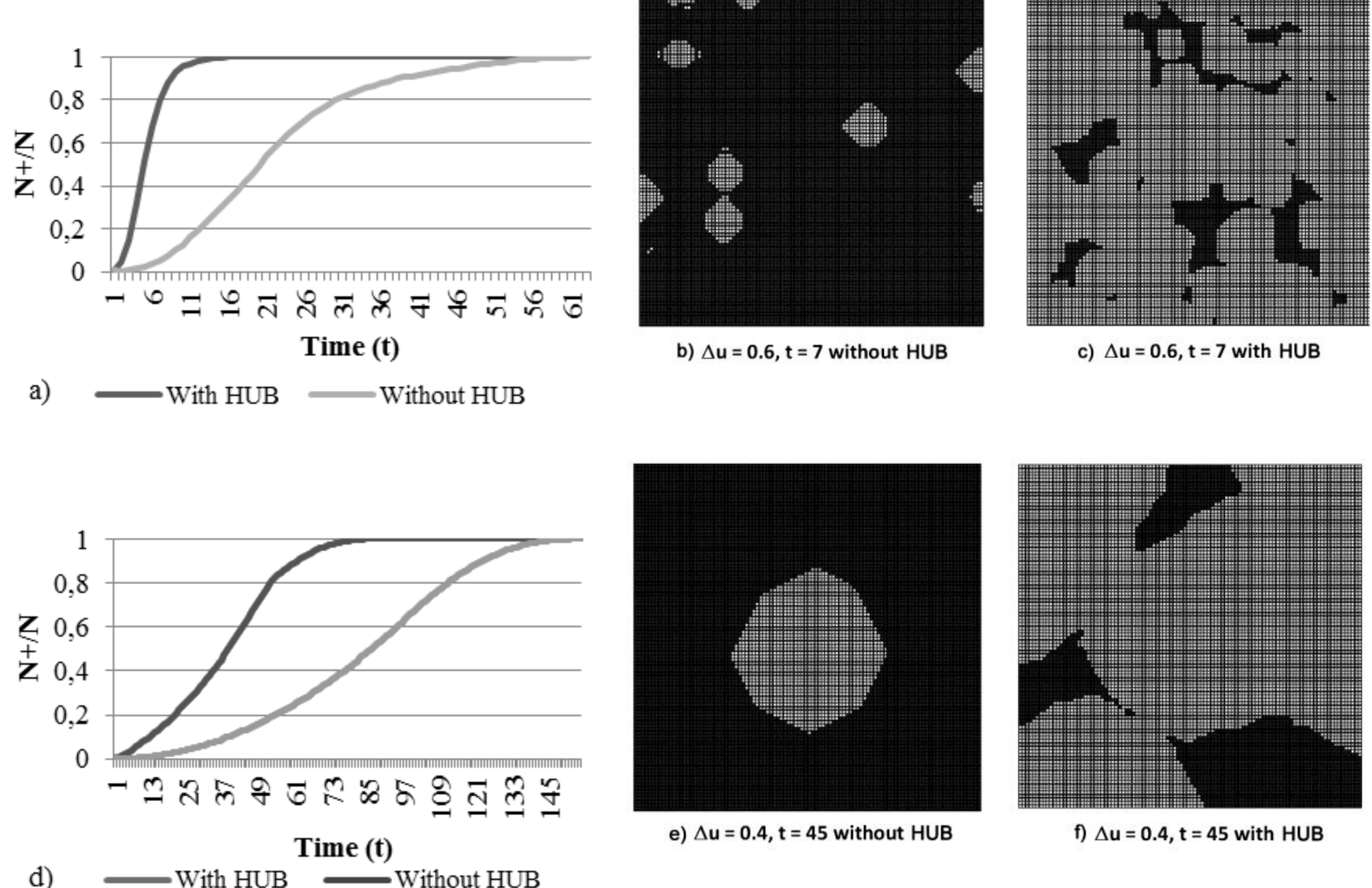

Figure 4: Comparison of adoption process for $\Delta u = 0.6$ (top figures), $\Delta u = 0.4$ (bottom figures) and rewiring probabilities $p = 0$ (without HUB) and $p = 0.1$ (with HUB). For $\Delta u = 0.6$, we use a uniform initial distribution of early adopters, whereas for $\Delta u = 0.4$ we use a distribution with $\sigma = 29$. The left chartsshow a comparison of the adoption curves for $\Delta u = 0.6$ (upper-left chart) and $\Delta u = 0.4$ (bottom-left chart). Figures a), b), c) and d) are snapshots of the adoption pattern for different combinations of $\Delta u$, p and time.

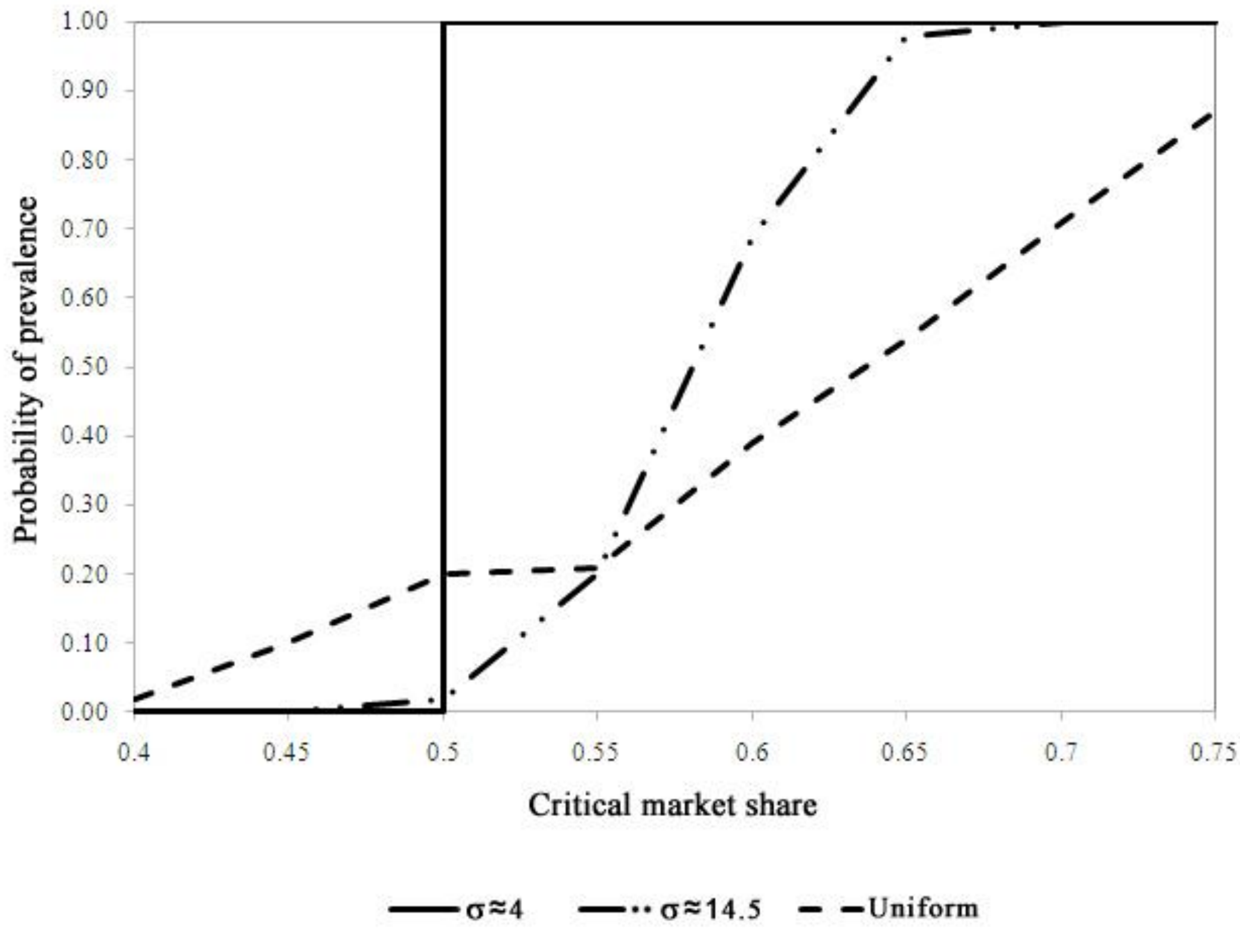

5a) Regular network

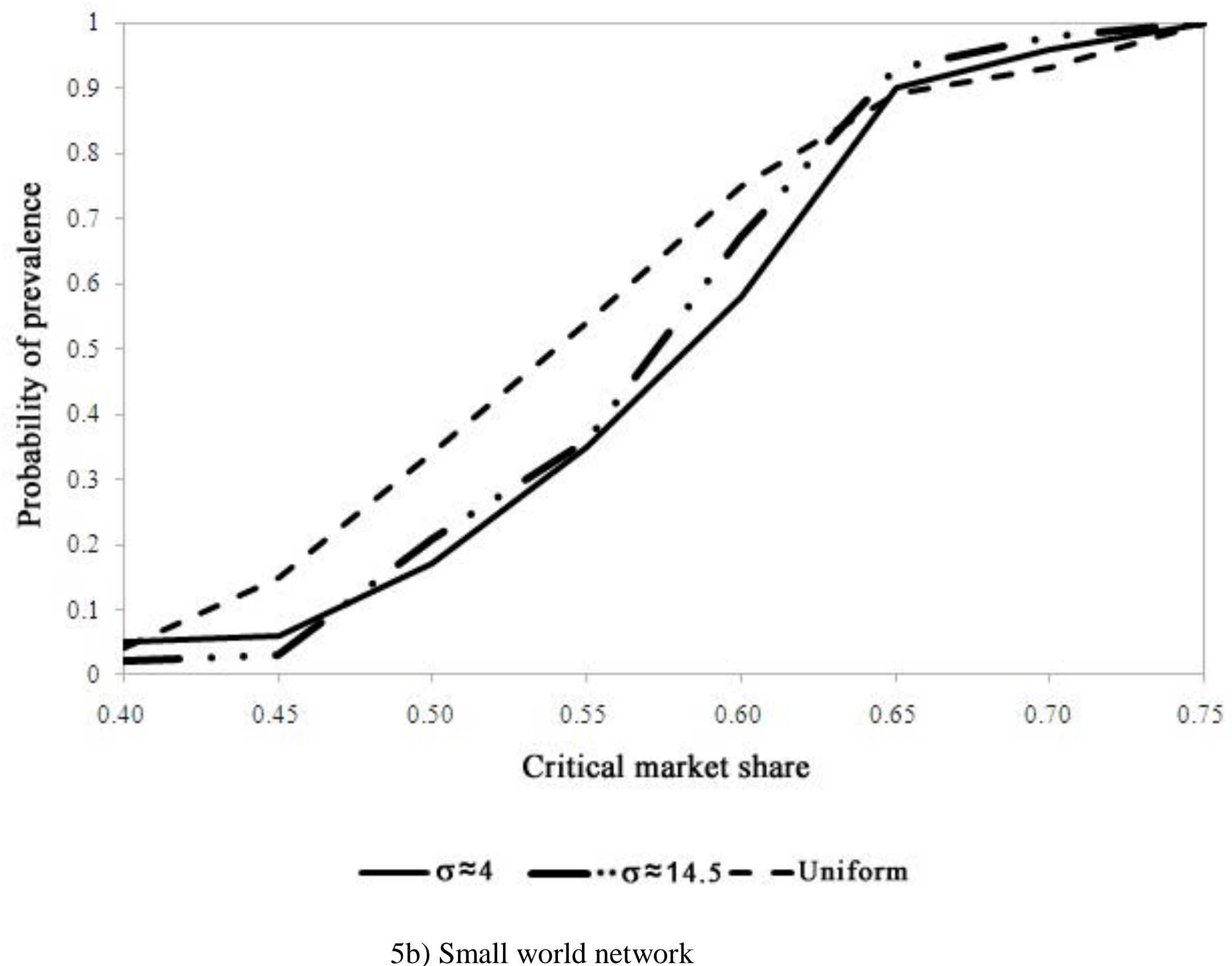

5b) Small world network

Figure 5: Probability of prevalence vs. the critical market share (CMS).